\documentclass{aa} 

\usepackage{graphicx}
\usepackage{txfonts}
\usepackage{natbib,twoopt}
\usepackage{amsmath}
\usepackage{multirow}
\usepackage[T1]{fontenc}
\usepackage[switch]{lineno}

\usepackage{multicol}

\begin{document}

    \title{Metal-silicate mixing in planetesimal collisions}

    \author{Kang Shuai\inst{1}\fnmsep\inst{2}
        \and
        Christoph M. Sch\"{a}fer\inst{2}
        \and
        Christoph Burger\inst{2}
        \and
        Hejiu Hui\inst{1}\fnmsep\inst{3}\fnmsep\inst{4}
        }

    \institute
        {State Key Laboratory for Mineral Deposits Research \& Lunar and Planetary Science Institute, School of Earth Sciences and Engineering, Nanjing University, Nanjing,
        210023, China\\
        \email{hhui@nju.edu.cn}
        \and
            Institut f\"{u}r Astronomie und Astrophysik, Eberhard Karls Universit\"{a}t T\"{u}bingen, Auf der Morgenstelle 10, 72076 T\"{u}bingen, Germany\\
            \email{ch.schaefer@uni-tuebingen.de}
        \and
            CAS Center for Excellence in Comparative Planetology, Hefei, 230026, China
        \and
            CAS Key Laboratory of Earth and Planetary Physics, Institute of Geology and Geophysics, Chinese Academy of Sciences, Beijing, 100029, China
            }
    \date{}
    \abstract
    {}
    {Impacts between differentiated planetesimals are ubiquitous in protoplanetary discs and may mix materials from the core, mantle, and crust of planetesimals, thus forming stony-iron meteorites. The surface composition of the asteroid (16) Psyche represents a mixture of metal and non-metal components. However, the velocities, angles, and outcome regimes of impacts that mixed metal and silicate from different layers of planetesimals are debated. Our aim is to investigate the impacts between planetesimals that can mix large amounts of metal and silicate, and the mechanism of stony-iron meteorite formation.}
    {We used smooth particle hydrodynamics to simulate the impacts between differentiated planetesimals with various initial conditions that span different outcome regimes. In our simulations, the material strength was included and the effects of the states of planetesimal cores were studied. Using a statistical approach, we quantitatively analysed the distributions of metal and silicate after impacts.}
    {Our simulations modelled the mass, depth, and sources of the metal-silicate mixture in different impact conditions. Our results suggest that the molten cores in planetesimals could facilitate mixing of metal and silicate. Large amounts of the metal-silicate mixture could be produced by low-energy accretional impacts and high-energy erosive impacts in the largest impact remnant, and by hit-and-run and erosive impacts in the second-largest impact remnant. After impact, most of the metal-silicate mixture was buried at depth, consistent with the low cooling rates of stony-iron meteorites. Our results indicate that mesosiderites potentially formed in an erosive impact, while pallasites potentially formed in an accretional or hit-and-run impact. The mixing of metal and non-metal components on Psyche may also be the result of impacts.}
    {}
    
    \titlerunning{Metal-silicate mixing during planetesimal collisions}
    \authorrunning{K. Shuai et al.}
    \maketitle

\section{Introduction}
    Planetesimals that were heated by the decay of short-lived radioactive nuclides could have experienced melting and differentiated into silicate mantles and metallic cores. The planetesimal differentiation and collisions between these differentiated bodies are ubiquitous in protoplanetary discs \citep{Bonsor2020, Bonsor2023}. The differentiated planetesimals in our Solar System are sampled by achondrites, iron meteorites, and stony-iron meteorites, which represent silicate, metal, and metal-silicate mixture in the planetesimals \citep{Greenwood2020}. These meteorites have survived many dynamical and geological processes, recording key information about planetesimal formation and evolution, and evolution of the protoplanetary disc \citep{Greenwood2020, Benedix2014, Hunt2022}. However, the origin of stony-iron meteorites is debated. Although there is general agreement that the silicate in pallasites and mesosiderites, the two main groups of stony-iron meteorites, originated respectively from the mantles and the crusts of planetesimals  \citep{Benedix2014}, the exact mechanism that mixed silicate with the core-derived metal remains debated \citep{Haba2019, IanniniLelarge2022, Kruijer2022, Sugiura2022, Windmill2022}.
    
    Pallasites are mainly composed of subequal amounts of olivine crystals and iron-nickel metal \citep{Benedix2014}, and were proposed to have formed at the core-mantle boundary of  differentiated planetesimals \citep{Wood1978, Wasson2003}. However, the core-mantle boundary origin of pallasites is difficult to reconcile with the diverse cooling rates and rare earth element data of the main group pallasites. Samples from the core-mantle boundary should have indistinguishable cooling rates \citep{Yang2010} and the rare earth element concentration and pattern in phosphates suggest that they are late-crystallizing phases in the mantle, which cannot crystallize at the base of mantle \citep{Davis1991, Hsu2003}. Impacts that tore apart a differentiated planetesimal were proposed to have produced the typical main group pallasites \citep{Yang2010}. On the other hand, the paleomagnetic records in the main group pallasites suggest that they resided in a body with an intact core that had dynamo activity \citep{Tarduno2012}. Therefore, pallasites were proposed to have formed in an accretional impact, in which the liquid metal from the core of a differentiated projectile was injected into the mantle of a target with a molten core \citep{Tarduno2012, Windmill2022}. Alternatively, intrusive ferrovolcanism could also mix the materials from the mantle and core of a planetesimal to form pallasites \citep{Johnson2020}. Unlike pallasites, the silicate in mesosiderites has basaltic and gabbroic textures, which is considered to be derived from the crust of a differentiated planetesimal \citep{Benedix2014}. Different scenarios of planetesimal impacts that could have mixed crust- and core-derived materials to form mesosiderites have been proposed, including low-velocity ($\leq$1 km s$^{-1}$) accretional impacts \citep{Wasson1985, IanniniLelarge2022} and high-velocity ($\sim$5 km s$^{-1}$, >10 times the mutual escape velocity of planetesimals at $\sim$200 km scale) hit-and-run or near-catastrophic impacts \citep{Scott2001, Haba2019, Sugiura2022}.
    
    The key parameters in the impact-induced models of stony-iron meteorite formation, such as impact velocity, impact angle, and the properties of the impacting planetesimals, can be investigated using numerical simulations. Early low-resolution smooth particle hydrodynamics (SPH) simulations revealed that near-catastrophic impacts can excavate core material and launch it on suborbital trajectories, which may reaccrete and mix with crustal material to form mesosiderites \citep{Scott2001}. Recent SPH simulations that neglected material strength reproduced the impact-induced mixing of the core and crust material on the surface of a Vesta-like body \citep{Sugiura2022}. However, mesosiderites record low metallographic cooling rates ($<$ 1 K Myr$^{-1}$), indicating that they formed deep within their parent body \citep{Haack1996}. The cooling rates of pallasites also indicate that they formed at depths of tens of kilometres  \citep{Bryson2015}. Metal-silicate mixing in the depth of planetesimals has not been modelled in previous studies. In addition, impacts could be transformative for the projectile (the smaller of the two impactors) \citep{Asphaug2014,Cambioni2021}, and the metal-silicate mixing in the surviving projectile has not been studied by numerical simulations. The metal-rich asteroid Psyche is likely a mixture of metal and non-metal components \citep{Elkins-Tanton2022}, which could be the remnant of mantle-stripping impacts \citep{Asphaug2014}.
    
    In this study we modelled the impact-induced mixing of metal and silicate and the burial of metal-silicate mixture in planetesimals using SPH simulations. Impacts that span different outcome regimes were simulated, including accretion, hit-and-run, erosion, and super-catastrophic disruption. We analysed the distribution of metal and silicate in the largest and the  second-largest remnants after impact to identify the metal-silicate mixture. The mass, depth, and sources of the metal-silicate mixture in different impact conditions were modelled, providing insights into the mechanism of stony-iron meteorite formation and the structure of asteroids in the Solar System.

\section{Methods}
    We modelled the impacts between differentiated planetesimals by three-dimensional SPH simulations with the \texttt{miluphcuda} code \citep{Schaefer2016, Schaefer2020}. Impacts spanning different outcome regimes were simulated, with different impact angles ($\theta$), impact velocities ($v$), projectile-to-target mass ratios ($\gamma$), and thus different energies. The impact angle was defined as the angle between the line connecting the centers of the two bodies and the projectile velocity vector at the time of first contact, with oblique impacts having high impact angles. The reduced mass kinetic energy, $Q_{\rm R}=0.5{\gamma v^2}/{(1+\gamma)^2}$ \citep{Stewart2009}, was used, with higher impact velocity and higher projectile-to-target mass ratio ($\gamma \leq 1$) corresponding to higher impact energy. In our simulations the material strength was considered, which has substantial effects on the outcome of collisions between planetesimals at $\lesssim$1000 km scale \citep{Jutzi2015}. Different states (solid or liquid) of the cores of impacting bodies were used to investigate the effects of the core's state on the metal-silicate mixing process. Using Moran's $I$ statistics \citep{moran1950}, we analysed the distribution of metal and silicate in the largest and the second-largest remnants after impact to identify the metal-silicate mixture that could be the source of stony-iron meteorites. The mass of metal-silicate mixture and its depth buried in the impact remnant were modelled.

    \subsection{Physical model}
        The Tillotson equation of state \citep{Tillotson1962} was used for both metal and silicate. Additional simulations for a few sets of initial conditions were conducted using the ANEOS equation of state \citep{Thompson1972, Melosh2007} to compare the effects of different equations of state. The SPH code includes self-gravity and is capable of modelling solid-body physics. A pressure-dependent yield criterion was used to model the plastic behaviour of all materials, following previous studies \citep{Collins2004, Jutzi2015}. The yield strength of intact materials is defined as
        \begin{linenomath}
            \begin{equation}
                \label{eq:1}
                Y_{\rm P} = c + \frac{\mu p}{1 + \mu p / (Y_{\rm M} - c)},
            \end{equation}
        \end{linenomath}
        where $c$ is the cohesion, $Y_{\rm M}$ is the shear strength at infinite pressure, and $\mu$ is the coefficient of internal friction.
        
        We used the Grady-Kipp fragmentation model for solid silicate, which simulates the activation and growth of flaws in a brittle material \citep{Benz1995}. The model parameters for the Weibull distribution of flaw activation thresholds ($m$ and $k$) are listed in Table~\ref{table:B1}. In the Grady-Kipp fragmentation model, a scalar damage variable $D$ is introduced to represent the reduction in strength under tensile loading with $0 \leq D \leq 1$, where $D = 0$ corresponds to an undamaged intact material and $D = 1$ to a fully damaged (fragmented) material. The yield strength of a totally damaged, granular material that has $D = 1$ is given by
        \begin{linenomath}
            \begin{equation}
                \label{eq:2}
                Y_{\rm D} = c_{\rm D} + \mu _{\rm D} p,
            \end{equation}
        \end{linenomath}
        where $c_{\rm D}$ and $\mu_{\rm D}$ represent the cohesion and the coefficient of internal friction for the totally damaged material, and $Y_{\rm D}$ is no greater than $Y_{\rm P}$. For the partially damaged material, the interpolation between $Y_{\rm P}$ and $Y_{\rm D}$ was used:
        \begin{linenomath}
            \begin{equation}
                \label{eq:3}
                Y = (1 - D) Y_{\rm P} + D Y_{\rm D}.
            \end{equation}
        \end{linenomath}
        The brittle-ductile transition temperature for iron meteorites under impact conditions is very low ($\sim$200 K) \citep{Johnson1974}. Therefore, the metal that made up cores of planetesimals would exhibit ductile behaviour during the formation of stony-iron meteorites. We did not use the Grady-Kipp fragmentation model for solid metal, and $Y=Y_{\rm P}$ was used.
        
        The specific melt energy $E_{\rm melt}$ was used to reduce the yield strength with increasing internal energy \citep{Benz1999, Jutzi2015}
        \begin{linenomath}
            \begin{equation}
                \label{eq:4}
                Y_{\rm melt}= \left \{
                \begin{aligned}
                    &Y\left(1-\frac{E}{E_{\rm melt}}\right) & E<E_{\rm melt}& \\
                    &0 & E\geq E_{\rm melt}&
                \end{aligned},
                \right.
            \end{equation}
        \end{linenomath}
        where $Y_{\rm melt}$ is the energy-dependent yield strength used in our simulations, and $E$ is the specific internal energy. If one particle exceeds the yield strength, the deviatoric stress is reduced with the limiting factor
        \begin{linenomath}
            \begin{equation}
                \label{eq:5}
                f_{\rm Y} = \min \left[Y_{\rm melt} / \sqrt{J_2},~1\right],
            \end{equation}
        \end{linenomath}
        where $J_2$ is the second invariant of the deviatoric stress tensor. The material strength parameters for solid metal and solid silicate are listed in Table~\ref{table:B1}.
        
        It was suggested that metal in mesosiderites and pallasites were largely molten when mixed with silicate \citep{Scott1977, Hassanzadeh1990}. Paleomagnetic evidence suggested that, during the formation and cooling of pallasites, the pallasite parent body had a molten core \citep{Tarduno2012}, which took hundreds of million years to solidify \citep{Bryson2015}. Therefore, we conducted impact simulations both for bodies with solid cores and for bodies with molten cores, using the same impact velocities, impact angles, and projectile-to-target ratios. The yield strength and the shear modulus of liquid metal were set to zero. The bulk modulus and the reference density of liquid metal are 109.7 GPa and 7019 kg m$^{-3}$ \citep{Anderson1994}. In our simulations, the planetesimals are assumed to be completely differentiated. The porosity could have reached zero almost in the entire body due to sintering and melting \citep{Neumann2012}. Therefore, porosity was neglected in our simulations. In addition, we performed purely hydrodynamic simulations to investigate the effects of material strength on planetesimal collisions.
    
    \subsection{Initial conditions}
        We simulated the impacts between differentiated planetesimals using various initial conditions to cover different outcome regimes. The Vestan origin of mesosiderites is still under debate \citep{Haba2019, IanniniLelarge2022}. Nevertheless, the estimated radius of the mesosiderite parent body (200$-$400 km) \citep{Haack1996} is consistent with the radius of Vesta ($\sim$270 km). Furthermore, the radius of Vesta is also comparable to the estimated radius of the pallasite parent body ($\sim$200 km) \citep{Tarduno2012}. In addition, the core fraction of a differentiated body is controlled by its oxygen fugacity (before disruption or mantle stripping), and the oxygen fugacities of Vesta and the pallasite parent body are similar \citep{Righter2020}. Therefore, we fixed the radius of target as 270 km and the radius of the target's core as 120 km in the simulations. We note that the pallasite parent body could have a large core of >200 km and a thin mantle of <70 km \citep{Nichols2021}. We discuss the effects of different metal-to-silicate ratios in section \ref{section4.3}. We used a semi-analytical approach to compute the relaxed internal structures of the pre-impact bodies; this approach  is capable of generating hydrostatic profiles that are very close to equilibrium \citep{Burger2018}. The mass of the target is $2.7 \times 10^{20}$ kg. Based on the core size, the target was composed of 21 wt.\% metal and 79 wt.\% silicate, and the same mass fractions of metal and silicate were used for the projectile. We used the projectile-to-target mass ratios ($\gamma$) of 0.1, 0.2, 0.5, and 1. The impact velocities ($v$) of 0.5 km s$^{-1}$, 1 km s$^{-1}$, 3 km s$^{-1}$, and 5 km s$^{-1}$ and the impact angles ($\theta$) of 0$^\circ$, 15$^\circ$, 30$^\circ$, 45$^\circ$, and 60$^\circ$ were used to provide good coverage for the impacts that can mix metal and silicate in planetesimals. The impact velocities varied from 1.4 to 15.9 times the mutual escape velocity. All simulations were run for a model time of 24 hours after impact to allow the reaccretion of most impact debris, the same as in previous simulations \citep{Rufu2017, Carter2018}. In most simulations, a standard resolution was used, with the target consisting of $10^5$ particles and the projectile consisting of $10^4-10^5$ particles.  The mass per particle in the projectile was the same as that in the target. In total, we ran 160 simulations in the standard resolution with material strength. Resolution tests were performed with the target consisting of $2\times10^5$, $5\times10^5$, and $10^6$ particles for three sets of initial conditions.

    \subsection{Identification of metal-silicate mixture}
        The two largest remnants after the impact in each simulation were analysed. The other impact remnants were too small to match the parent bodies of stony-iron meteorites. In the largest and the second-largest impact remnants, each particle and its nearest particles (50 particles in total) were analysed, which determined whether metal and silicate were mixed around this particle. The global Moran's $I$ statistics was used to measure the extent of mixing \citep{moran1950}
        \begin{linenomath}
            \begin{equation}
                \label{eq:6}
                I=\frac{N}{W} \frac{\Sigma^N_{i=1} \Sigma^N_{j=1} w_{i,j}(x_i-\bar{x})(x_j-\bar{x})}{\Sigma^N_{i=1}(x_i-\bar{x})^2},
            \end{equation}
        \end{linenomath}
        where $N=50$ is the number of particles in the region where Moran's $I$ is measured; $x$ represents the material of the particle with $x=0$ for metal and $x=1$ for silicate; $\bar{x}$ is the mean of $x$ for all measured particles; and $w_{i,j}$ is the spatial weight determined by whether particle $j$ is the neighbour of particle $i$. We set $w_{i,j}=1$ if particle $j$ is the closest particle to particle $i$, and $w_{i,j}=0$ in the other cases. $W$ is the sum of $w_{i,j}$:
        \begin{linenomath}
            \begin{equation}
                \label{eq:7}
                W=\Sigma^N_{i=1} \Sigma^N_{j=1} w_{i,j}.
            \end{equation}
        \end{linenomath}
        The calculated $I$ for each particle measures the spatial autocorrelation of metal and silicate in the region around this particle. If metal and silicate are extensively mixed, the particles of these two materials have negative spatial autocorrelation and $I$ approaches $-1$, while $I$ is close to 1 in the regions such as the core-mantle boundaries of pre-impact bodies where metal and silicate are clustered. The expected value of Moran's $I$ under the null hypothesis of no spatial autocorrelation is $E(I)$,
        \begin{linenomath}
            \begin{equation}
                \label{eq:8}
                E(I)=-\frac{1}{N-1},
            \end{equation}
        \end{linenomath}
        which represents the random distribution of metal and silicate. In addition, the metal-silicate mixture that is possibly the source of stony-iron meteorites should have large amounts of both metal and silicate. Therefore, metal-silicate mixture is defined as the regions that have $I<E(I)$ and have both a metal mass fraction and a silicate mass fraction within the range  0.3$-$0.7. A particle is regarded as metal-silicate mixture if its nearest 50 particles meet these criteria. The depth of a particle in the impact remnant was calculated as the shortest distance between the particle and the surface of this remnant.

    \subsection{Classification of impact regimes}
        The impact outcomes are classified into four regimes mainly according to their accretion efficiency $\xi$ \citep{Asphaug2009,Leinhardt2012}:
        \begin{linenomath}
            \begin{equation}
                \label{eq:9}
                \xi=(M_{\rm l}-M_{\rm t})/M_{\rm p}.
            \end{equation}
        \end{linenomath}
        Here $M_{\rm l}$ is the mass of the largest impact remnant, $M_{\rm t}$ is the mass of the target, and $M_{\rm p}$ is the mass of the projectile. The impacts with $\xi>0.2$ are classified as accretional impacts, in which a large fraction of mass from the projectile accreted onto the target. The accretional impacts generally have low impact energies and impact angles, which led to a large amount of mass bounded to the largest impact remnants. The hit-and-run impacts have $-0.2\leq\xi\leq0.2$, with the target mass almost unchanged during the impacts. The projectiles grazed the targets at a large impact angle in these hit-and-run impacts. The impacts with $M_{\rm l}/(M_{\rm t}+M_{\rm p})<0.1$ are super-catastrophic impacts \citep{Leinhardt2012} that have high impact energies and low impact angles. Other high-energy impacts with $\xi<-0.2$ and $M_{\rm l}/(M_{\rm t}+M_{\rm p})\geq0.1$ are classified as erosive impacts. In an erosive impact, the projectile is separated from the target after impact, as in the hit-and-run impacts, but a considerable fraction of the target is lost. We note that different criteria of impact regime classification were used in previous studies, and thus the same impacts could be classified as different regimes using different criteria (Appendix \ref{appendix:c}).

    \begin{figure*}
    \centering
    \includegraphics[width=15 cm]{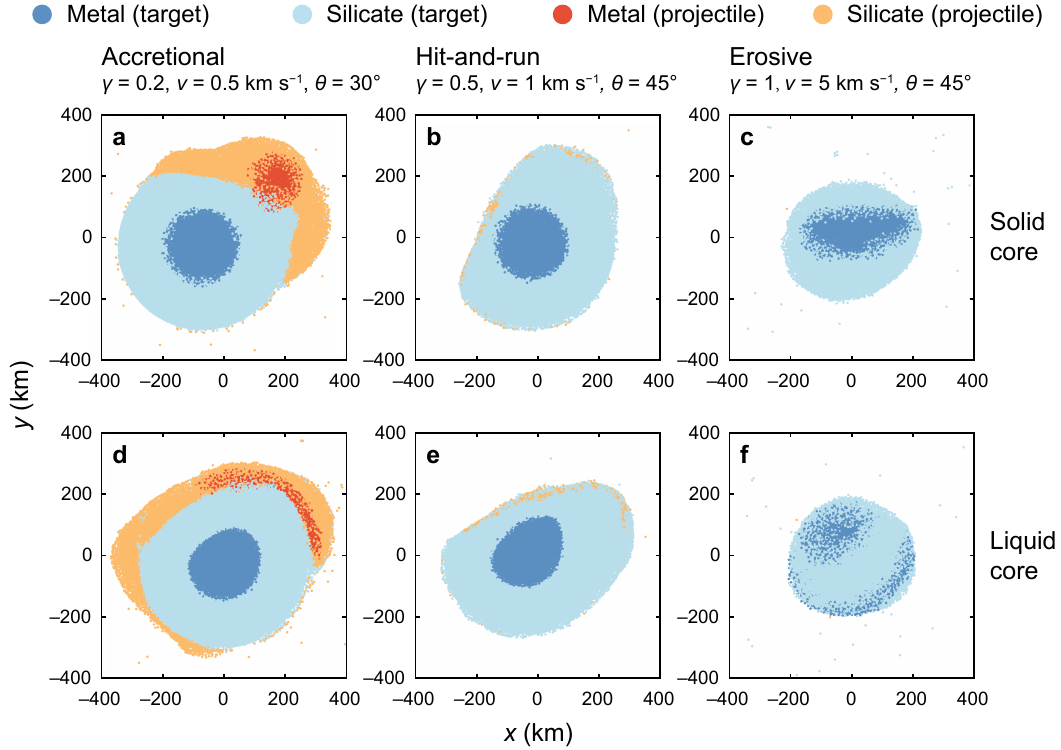}
    \caption{Distribution of metal and silicate at the end of simulations. The cross-sections of the largest impact remnants in different impact regimes are shown, which are in the plane of the projectile trajectory. The materials and sources of the particles are colour-coded (see legend): the projectile's metal and silicate are red and yellow, and the target's metal and silicate are dark blue and light blue. The impacts are accretional impacts that have a projectile-to-target ratio ($\gamma$) of 0.2; impact velocity ($v$) of 0.5 km s$^{-1}$ (1.6 times the mutual escape velocity, $v_{\rm esc}$); impact angle ($\theta$) of 30$^\circ$  (\textbf{a} and \textbf{d}); hit-and-run impacts with $\gamma=0.5$, $v=1.0$ km s$^{-1}$, $v/v_{\rm esc}=3.0$, $\theta=45^\circ$ (\textbf{b} and \textbf{e}); and erosive impacts with $\gamma=1.0$, $v=5.0$ km s$^{-1}$, $v/v_{\rm esc}=13.7$, $\theta=45^\circ$ (\textbf{c} and \textbf{f}). The cores of the projectiles and the targets were solid (\textbf{a-c}) or liquid (\textbf{d-f}).
    }
        \label{fig:1}
    \end{figure*}

    \begin{figure}
    \includegraphics[width=\hsize]{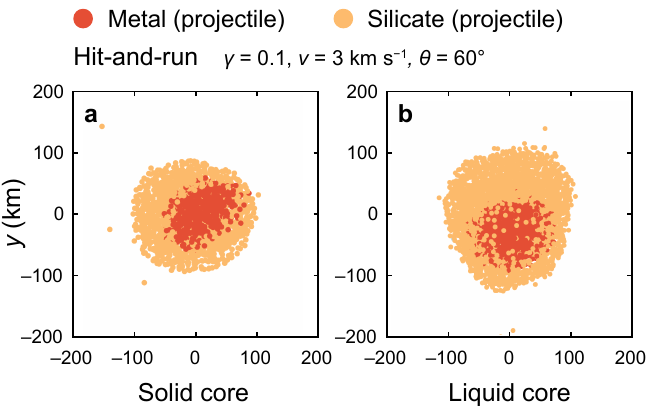}
    \caption{Distributions of metal and silicate in the second-largest impact remnants at the end of simulations. The cross-sections are in the plane of the projectile trajectory. All particles originate from the projectile. The metal and silicate particles are red and yellow, respectively. The impacts are hit-and-run impacts with $\gamma=0.1$, $v=3.0$ km s$^{-1}$, $v/v_{\rm esc}=9.5$, and $\theta=60^\circ$. The cores of the projectile and the target were solid (\textbf{a}) or liquid (\textbf{b}).
  }
        \label{fig:2}
    \end{figure}

    \begin{figure*}
    \centering
    \includegraphics[width=15 cm]{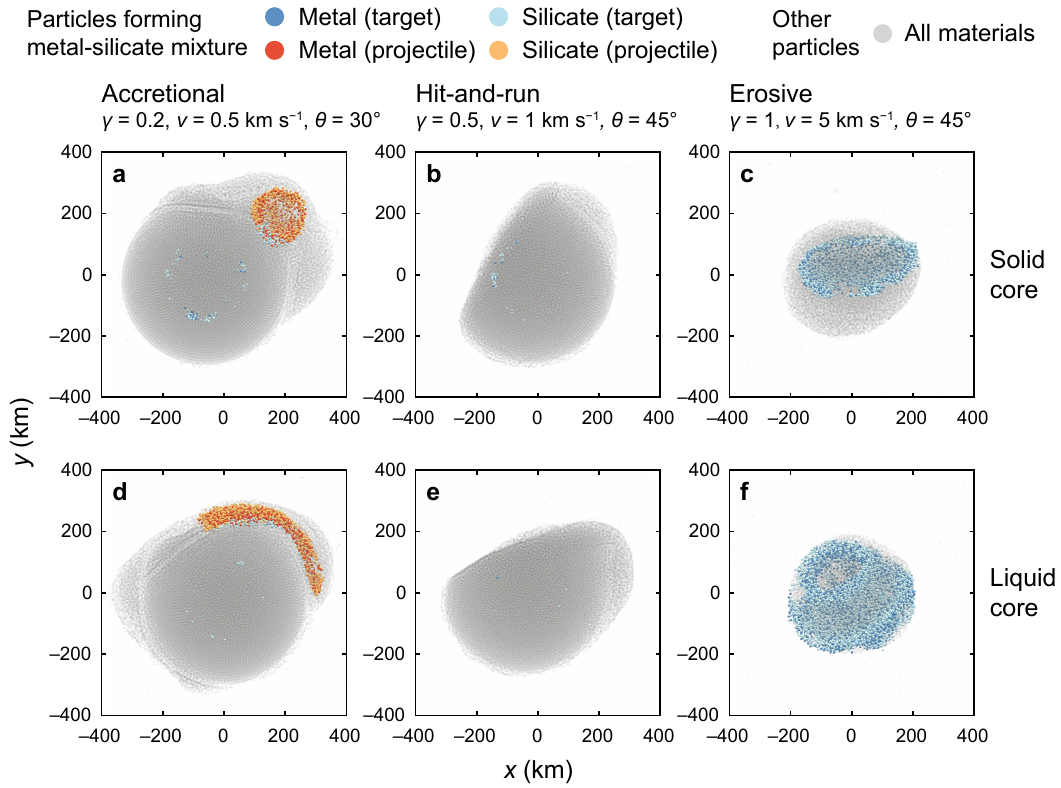}
    \caption{Distribution of metal-silicate mixture at the end of simulations. Shown are the cross-sections of the largest impact remnants in different impact regimes, which are in the plane of the projectile trajectory. The materials and sources of the particles forming metal-silicate mixture are colour-coded (see legend): the projectile's metal is red, the projectile's silicate is yellow, the target's metal is dark blue, and the target's silicate is light blue. Each coloured particle was identified by calculating the metal mass fraction (within 0.3$-$0.7) and Moran's $I$ ($I<E(I)$) of its nearest 50 particles. Particles that did not form metal-silicate mixture are in transparent gray. The initial conditions of the impacts are the same as those in Fig.~\ref{fig:1}.
    }
        \label{fig:3}
    \end{figure*}

\section{Results}

    The distributions of metal and silicate are significantly different after impacts with different initial conditions (Figs.~\ref{fig:1}, ~\ref{fig:A1}, and ~\ref{fig:A2}). In the largest impact remnant, metal and silicate were extensively mixed in accretional impacts and erosive impacts (Fig.~\ref{fig:1}a, c, d, and f), while metal and silicate remained separated in hit-and-run impacts (Fig.~\ref{fig:1}b and e). On the other hand, metal and silicate could be mixed in the second-largest remnant after hit-and-run and erosive impacts (Fig.~\ref{fig:2}). The projectiles were eroded and survived as the second-largest remnant. Regions with a metal mass fraction within 0.3$-$0.7 and $I<E(I)$ indicative of negative spatial autocorrelation of metal and silicate were identified (Fig.~\ref{fig:3}). The smoothed particles representing these regions were defined as metal-silicate mixture. Regions where metal and silicate are separated, such as the sharp core-mantle boundary (e.g. Fig.~\ref{fig:1}e and Fig.~\ref{fig:3}e), were not included, showing that the Moran's $I$ is sensitive to the degree of mixing. For each simulation, the masses of metal-silicate mixture in the largest and the second-largest impact remnants were calculated. These masses may not be identical to the masses of iron-stony meteorites in the remnants because our large-scale simulations cannot precisely reproduce the mixing of centimetre-sized silicate and metal grains. However, the identified metal-silicate mixture in our simulation results can be considered the potential source region of iron-stony meteorites, where metal and silicate from different layers of the impacting planetesimals have been mixed. Further studies of subsequent collisions and exhumation are required to prove that iron-stony meteorites can be exhumed from these regions without being remelted.
    
    The simulations using the ANEOS equation of state \citep{Thompson1972, Melosh2007} have post-impact distributions of materials (Fig.~\ref{fig:A1}) similar to those in simulations using the Tillotson equation of state \citep{Tillotson1962} (Fig.~\ref{fig:1}). The simulations using different equations of state also produced similar masses of metal-silicate mixture (Fig.~\ref{fig:A3}). Therefore, different equations of state do not affect the post-impact distribution of materials significantly, consistent with the previous study on larger-scale impacts \citep{Emsenhuber2018}.
    
    It has been demonstrated that spurious dissipation of subsonic turbulence results in inaccurate material mixing in SPH simulations, and that higher resolution in general leads to more accurate mixing \citep{Agertz2007, Deng2019, Gabriel2023}. In our results, for impacts with the same parameters (impact angles, impact velocities, projectile-to-target ratios, and core states of planetesimals), the variations in masses of metal-silicate mixture between different resolutions are much larger than those variations between different equations of state (Fig.~\ref{fig:A3}). The masses of metal-silicate mixture do not change monotonically with resolution, resulting in uncertainties in our results. However, the distribution of materials at the end of simulations with a resolution ten times the standard resolution (Fig.~\ref{fig:A2}) are similar to those in the standard-resolution simulations (Fig.~\ref{fig:1}). In addition, different impact regimes can be discerned based on the masses of metal-silicate mixture (Fig.~\ref{fig:A3}). The uncertainty of the metal-silicate mixture mass can be estimated using the dispersion (2$\sigma$) of the metal-silicate mixture masses in impacts with the same impact parameters and different resolutions. For impacts that produced high masses of metal-silicate mixture in the largest impact remnant (the accretional and erosive impacts), the  2$\sigma$ is $\sim10-40$\%. Only metal-silicate mixture with very low masses has high relative uncertainties because the mixture is represented by a small number of particles. However, the absolute uncertainties of these low masses are of the order of 10$^{17}$ kg, which do not change the results significantly. Therefore, our SPH simulations could provide constraints on masses of metal-silicate mixture after impacts in different outcome regimes. We also tested different numbers of particles for the Moran's $I$ analysis ($N$) by analysing the results of the same simulation with different $N$. The masses of metal-silicate converge as $N$ increases for $N \geq 50$ (Fig.~\ref{fig:A4}).

\section{Discussion}
    \subsection{Effects of material states on metal-silicate mixing}
        In our simulations the solid materials were simulated with material strength, while the materials without material strength could be considered the analogues of the molten metal and silicate. The results of purely hydrodynamic simulations (Fig.~\ref{fig:A5}) are distinct from those of the simulations using material strength (Fig.~\ref{fig:1}), especially for low-energy impacts. After a low-energy accretional impact with 0.5 km s$^{-1}$ velocity and a small projectile ($\gamma = 0.2$), metal from the projectile sank to the core-mantle boundary of the target in the purely hydrodynamic simulation (Fig.~\ref{fig:A5}a), whereas the projectile's metal remained at the surface of the target's mantle in the simulations using material strength (Fig.~\ref{fig:1}a and d). Our results are consistent with simulations of hit-and-run impacts and accretional impacts with inclusion of material strength \citep{Emsenhuber2024}. Radiogenic heating, cooling, and melting on a planetesimal depend on its size and formation time \citep{Elkins-Tanton2011, Lichtenberg2016}. A Vesta-sized planetesimal used as the projectile for $\gamma=1$ cases and the target in our simulations would only have a shallow magma ocean with a depth of up to a few tens of kilometres (compared with its $\sim$150 km thick silicate part), even if it formed early with high abundances of short-lived radionuclides \citep{Neumann2014}. It was suggested that the magma ocean of Vesta had a short lifetime of 2$-$3 Myr \citep{Schiller2011}, which could have cooled rapidly due to efficient removal and foundering of the crustal lid of the magma ocean \citep{Mandler2013}. Smaller planetesimals as the projectiles for $\gamma<1$ cases in our simulations would cool more rapidly \citep{Elkins-Tanton2011, Lichtenberg2016}. Furthermore, the nearly disruptive impacts on small bodies with diameters no greater than a few hundred kilometres cannot produce large-scale melting, and more energetic impacts could shatter the bodies instead of leaving intact molten bodies \citep{Keil1997}. Therefore, impacts between Vesta-size or smaller differentiated planetesimals with completely molten mantles would be rare in the Solar System. In addition, the metal mixed with silicate in a molten mantle could sink to the core over geological timescales \citep{Lichtenberg2023} and may not be excavated as stony-iron meteorites. We focus on the impacts between planetesimals with a solid silicate part and investigate the effect of core melting on the metal-silicate mixing process.
        
\begin{figure*}
    \sidecaption
    \includegraphics[width=12cm]{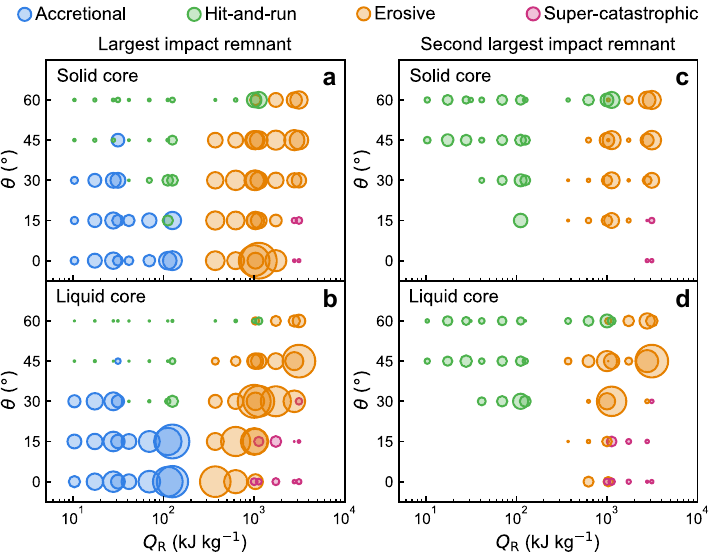}
    \caption{Mass of metal-silicate mixture in the largest and the second-largest impact remnants. Each circle corresponds to one impact simulation. The sizes of the circles are proportional to the mass of metal-silicate mixture produced in the impacts. The impacts with higher impact velocities and higher projectile-to-target ratios have higher impact energies ($Q_{\rm R}$). The outcome regimes of impacts are classified according to their accretion efficiency (see legend).
    }
        \label{fig:4}
    \end{figure*}

        In the impacts between planetesimals with molten cores, the liquid metal tended to spread in silicate, facilitating the mixing of metal and silicate (Fig.~\ref{fig:1}d and f; Fig.~\ref{fig:2}b). By contrast, the solid metal was torn apart with greater difficulty in impacts, leading to intact cores with a scrambled core-mantle boundary in the simulations with solid cores (Fig.~\ref{fig:1}a and c; Fig.~\ref{fig:2}a). Generally, the impacts between planetesimals with liquid cores produced higher masses of metal-silicate mixture than those with solid cores, especially for accretional impacts (Fig.~\ref{fig:4}). Only for high-energy ($Q_{\rm R}\sim10^3$ kJ kg$^{-1}$) head-on ($\theta=0^\circ$) impacts was more metal-silicate mixture produced in the simulations with solid cores. The strength of solid metal might have impeded the super-catastrophic disruption of the planetesimals, and more metal-silicate mixture was preserved in the largest impact remnant. However, head-on impacts between planetesimals were rare in the protoplanetary disc and the most frequent impact angle was around 45$^\circ$ \citep{shoemaker1962}. Therefore, the molten cores in planetesimals could facilitate the mixing of metal and silicate in most planetesimal collisions. Metal-silicate mixing would be easier at the early stage of the Solar System, when the cores of planetesimals were still molten. The facilitation of metal-silicate mixing by core melting is consistent with the mineralogical and paleomagnetic evidence that suggests molten cores in the parent bodies of stony-iron meteorites \citep{Scott1977, Hassanzadeh1990, Tarduno2012, Bryson2015}.
        
        In solid mantles of planetesimals, liquid metal could percolate through the mantle due to the density contrast between metal and silicate \citep{Ghanbarzadeh2017}. However, deformation experiments demonstrated that the formation of angular and fragmental pallasites requires rapid cooling after impact until reaching the Fe-Ni-S solidus within months to years, followed by slow conductive cooling \citep{Walte2020}. Liquid metal could have solidified in rapid cooling, which froze the metal-silicate mixture in the mantle. In addition, mineralogy of silicate in both pallasites and mesosiderites shows that the initial cooling rates at high temperature are orders of magnitude faster than the low-temperature metallographic cooling rates \citep{Ganguly1994, Ruzicka1994, Hsu2003}. The rapid cooling was proposed to result from impact rebound \citep{Walte2020}. Therefore, pallasites and mesosiderites that formed in impacts could have experienced rapid cooling and stayed in the mantle of their parent bodies. 

    \subsection{Metal-silicate mixing in different impact regimes}
\begin{figure}
    \centering
    \includegraphics{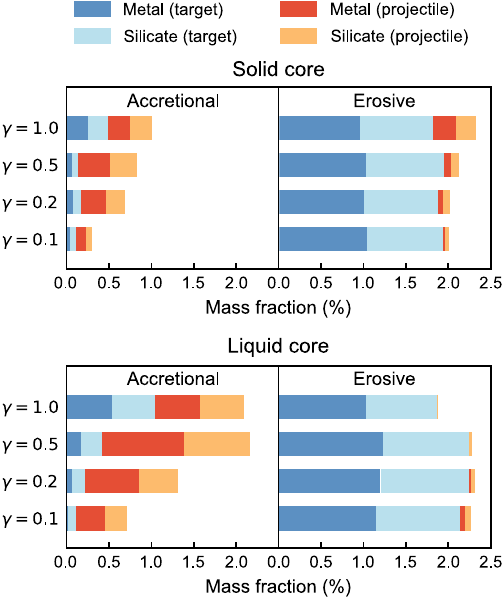}
    \caption{Average mass fraction of metal-silicate mixture in the largest impact remnant. The results of accretional or erosive impacts with different impact angles, impact velocities, and the same projectile-to-target ratios ($\gamma$) are averaged. The materials and sources of the particles composing the metal-silicate mixture are colour-coded (see legend).
    }
        \label{fig:5}
    \end{figure}

        Impacts in different outcome regimes produced different masses of metal-silicate mixture (Fig.~\ref{fig:4}). In the hit-and-run impacts with high impact angles, the target's core was unchanged or just deformed, with metal and silicate remaining separated (Fig.~\ref{fig:1}b and e). Most hit-and-run impacts produced little mass of metal-silicate mixture in the largest impact remnant (Fig.~\ref{fig:4}a and b). In the super-catastrophic impacts a large number of small fragments were produced, which would be ground down by successive collisions and depleted by radiation pressure \citep{Genda2015}. Therefore, the metal-silicate mixture produced in super-catastrophic impacts might be difficult to be preserved in the protoplanetary disc. The accretional impacts and erosive impacts mixed much more metal and silicate than the hit-and-run impacts and super-catastrophic impacts in the largest impact remnant. On average, the metal-silicate mixture produced in the accretional impacts and erosive impacts accounted for a small  percentage of the mass in the largest impact remnant (Fig.~\ref{fig:5}). On the other hand, hit-and-run impacts can be transformative for the projectile \citep[Fig.~\ref{fig:2};][]{Asphaug2014,Cambioni2021}, producing more metal-silicate mixture in the second-largest impact remnant than in the largest one  (Fig.~\ref{fig:4}c and d). In many erosive impacts, high masses of metal-silicate mixture were also produced in the second-largest remnant (Fig.~\ref{fig:4}c and d), whereas, in the accretional impacts no large remnant exists except for the largest one. The erosive impacts have $Q_{\rm R}> 200$ kJ kg$^{-1}$ and $v>1$ km s$^{-1}$, more energetic than the accretional impacts (Fig.~\ref{fig:4}). If another criterion based on the mass of the second-largest impact remnant is used to define impact regimes, some erosive impacts capable of producing high masses of metal-silicate mixture in the largest impact remnant can be classified as hit-and-run impacts, and almost exclusively all impacts that can mix metal and silicate in the second-largest impact remnant are classified as hit-and-run impacts (Appendix ~\ref{appendix:c}).
        
        We traced the sources of the metal and silicate particles composing metal-silicate mixture, which are different in the accretional and the erosive impacts (Fig.~\ref{fig:5}). In the accretional impacts, most of the metal and silicate in metal-silicate mixture originated from the projectile. The exceptions are the impacts between equal-mass bodies ($\gamma=1$). The accretional impacts with $\gamma=1$ have symmetric results with the metal-silicate mixture contributed equally by the target and the projectile. On the other hand, almost all the mixture of metal and silicate in the largest impact remnant were from the target in the erosive impacts. The second-largest impact remnant in hit-and-run and erosive impacts is the surviving projectile. Its metal-silicate mixture was formed by impact-driven erosion and originated from the projectile.

    \subsection{Buried depth of metal-silicate mixture} \label{section4.3}
\begin{figure}
    \centering
    \includegraphics{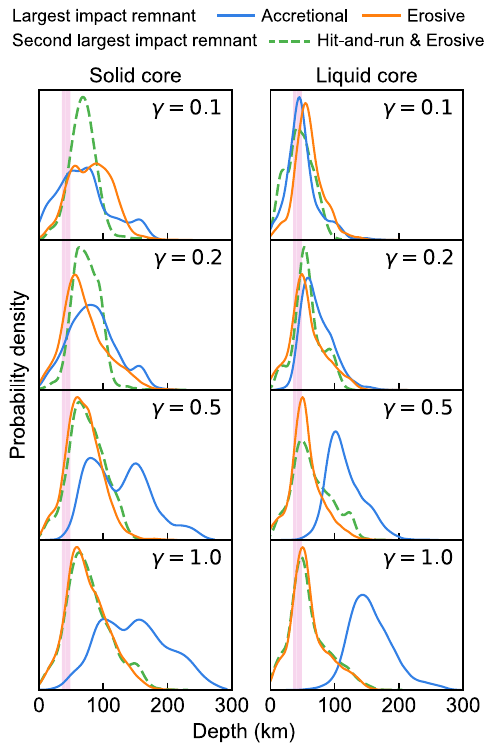}
    \caption{Depth distribution of metal-silicate mixture. The depth of a particle is the shortest distance between this particle and the surface of the impact remnant. The curves are the kernel density estimates for the depth of metal-silicate mixture produced by impacts with different projectile-to-target mass ratios ($\gamma$) and states of the planetesimal cores. The pink shaded bands show the depth of two pallasites in their parent body (38$\pm$2.5 km and 45$\pm$4 km) estimated using their cooling rates \citep{Bryson2015}.
    }
        \label{fig:6}
    \end{figure}

        The metal-silicate mixture in the impact remnants could be the potential source region of the stony-iron meteorites, where the materials from different layers of planetesimals were mixed. We analysed the depth of the particles composing metal-silicate mixture in the two largest impact remnants (Fig.~\ref{fig:6}). The depth was determined by measuring the shortest distance between the particle and the surface of the remnant. In all simulations, only a small fraction of the metal-silicate mixture resided on the surface of the two largest impact remnants. Most of the metal excavated by impacts was then covered by the reaccreted silicate, which was also observed in the high-resolution simulation in the previous study \citep{Sugiura2022}, consistent with the low cooling rates of stony-iron meteorites.
        
        The metal-silicate mixture was buried at different depths in different impact regimes (Fig.~\ref{fig:6}). After the erosive impacts, most of the metal-silicate mixture was buried at $\sim$50 km depth in the largest impact remnant, which is independent of the projectile mass. A similar buried depth was also observed for the metal-silicate mixture in the second-largest impact remnant, especially when the projectile-to-target mass ratio is close to 1. In the erosive impacts, a majority of the mixed metal and silicate in the largest impact remnant originated from the target (Fig.~\ref{fig:5}), which was buried at the same depth possibly due to the fixed size of the  target in our simulations. In the accretional impacts the depth of the metal-silicate mixture varies with $\gamma$. The metal and silicate mainly mixed and resided under the mantle of the projectile or in the mantles of the two equal-mass bodies when $\gamma=1$. As a result, the buried depth of the metal-silicate mixture in the accretional impacts increases with larger projectiles (higher $\gamma$), which  have thicker mantles.
        
        The metallographic cooling rate of pallasites indicates that they resided at a depth of $\sim$40 km \citep{Bryson2015}. This depth is consistent with the modelled depth of the metal-silicate mixture in the largest impact remnant in the accretional impacts with low $\gamma$ and all erosive impacts, and with the depth of metal-silicate mixture in the second-largest impact remnant in our simulations. We note that the pallasite parent body could have a non-chondritic metal-to-silicate ratio with a thin mantle, and the depth of pallasites could be different \citep{Nichols2021}. Although this thin-mantle model is not included in our initial conditions, it suggests that pallasites formed in the middle of the mantle, consistent with the previous studies \citep{Tarduno2012, Bryson2015} and with our results. The thin mantle could also affect the sources of the metal-silicate mixture in accretional impacts because the shallow core-mantle boundary could allow the target's core to be involved in mixing during impact. The lower metallographic cooling rate of mesosiderites indicates a deeper buried location \citep{Haack1996}. However, if the parent body of mesosiderites is Vesta or another intact body, only the metal-silicate mixture that can be excavated by later impacts can be ejected and reach Earth. The deepest material excavated by the two large impacts on Vesta came from $\sim$100 km depth \citep{Jutzi2013}, which could be the upper limit of the depth that mesosiderites resided.

    \subsection{Formation of stony-iron meteorites}
        Impacts were ubiquitous in the early Solar System, which caused disruption and mantle stripping of planetesimals as evidenced by the various groups of iron meteorites \citep{Benedix2014, Hunt2022}. Our simulations show that metal and silicate in planetesimals could be mixed by impacts, which could be the origin of stony-iron meteorites. On average, the two largest remnants after impacts contained a few weight per cent of metal-silicate mixture, consistent with the rarity of stony-iron meteorites in the meteorite collection ($<$1\% by number and $\sim$3\% by mass; \citeauthor{MetBulldatabase}). We note that the abundance of stony-iron meteorites in the meteorite collection is also controlled by other processes such as asteroid disruption and meteorite exhumation, and thus may not directly reflect the abundance of metal-silicate mixture in asteroids.
        
        Planetesimal collisions depend on the Solar System dynamics. Low-energy and high-energy impacts could be the major sources of the stony-iron meteorites that formed at different stages of the early Solar System. In the first several million years of the Solar System, the gas disc damped the eccentricity and inclination of planetesimals, which stabilized the disc and kept the impact velocities of planetesimals relatively low \citep{Kokubo2000, Morbidelli2012, Walsh2019}. After the dissipation of nebular gas, the dynamical excitation driven by the giant planets and the planetary embryos led to more energetic impacts between planetesimals \citep{Davison2013, Walsh2019}. \cite{Walsh2019} proposed that the average impact velocity between planetesimals was $\sim$0.46 km s$^{-1}$ at 0.7 Myr and 1 au where the gas damping effect was strong, and $\sim$2.3 km s$^{-1}$ at 18 Myr and 2 au where the gas had almost dissipated. The average impact velocity between main-belt asteroids is $\sim$5 km s$^{-1}$ and   collisions with velocities lower than 1 km s$^{-1}$ are very rare \citep{Bottke1994}. In our simulations, the impact velocities for accretional impacts are no greater than 1 km s$^{-1}$, comparable to the planetesimal collisions within a gas disc. The simulated erosive impacts have impact velocities of 3 km s$^{-1}$ and 5 km s$^{-1}$, which are consistent with   planetesimal collisions after the dissipation of nebular gas and   collisions in the main belt.
    
        \begin{figure}
        \centering
        \includegraphics{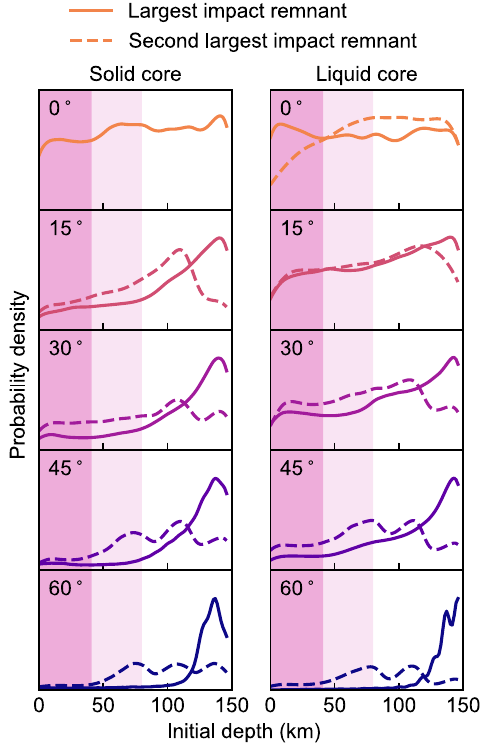}
        \caption{Initial depth distribution of the silicate mixed with metal in the erosive and hit-and-run impacts. The initial depth of a particle is its depth in the target or the projectile. The curves are the kernel density estimates for the initial depth of silicate that was mixed with metal in the erosive and hit-and-run impacts with different impact velocities, impact angles, and states of the planetesimal cores. The pink shaded bands show the thickness of Vesta's crust ($\sim$40 km and $\sim$80 km) estimated by previous studies \citep{Mandler2013, Clenet2014}.
        }
            \label{fig:7}
        \end{figure}
        
        Observations of circumstellar discs in young clusters suggest that disc lifetimes are typically a few million years \citep{Haisch2001}. The gas disc dissipation in our Solar System occurred at $\sim$4 Myr after Solar System formation, as suggested by records of the solar nebula magnetic field in meteorites \citep{Wang2017}. The Pd-Ag ages of iron meteorites show that they cooled rapidly at 7.8$-$11.7 Myr after Solar System formation, indicating disruptive impacts in an energetic inner Solar System, which probably resulted from the dissipation of nebular gas and the early instability of giant planets \citep{Hunt2022}.
        
        Radiometric dating shows that mesosiderites formed at 4525.39 $\pm$ 0.85 million years ago ($\sim$41.9 Myr after Solar System formation) \citep{Haba2019}, later than the gas disc dissipation, indicating that mesosiderites likely formed in an energetic impact. An energetic impact could be hit-and-run or erosive. We further analysed the initial depth (depth in the pre-impact bodies) of silicate particles composing the metal-silicate mixture in hit-and-run and erosive impacts (Fig.~\ref{fig:7}). In the largest impact remnant, the erosive impacts with lower impact angles mixed a larger amount of silicate from the crust of the target with metal. In the high-angle erosive impacts, almost no crust material was mixed with metal in the largest impact remnant because the crust of the target was preferentially eroded and the core tended to  mix with the mantle near the core-mantle boundary. In the second-largest impact remnant, most silicate in the  metal-silicate mixture originated deeper than 80 km. Therefore, mesosiderites, the mixture of crust and core materials, potentially formed in an erosive impact with a low impact angle. However, hit-and-run impacts and high-angle erosive impacts cannot be excluded. Our results indicate that, in erosive and hit-and-run impacts, the main source of metal-silicate mixture in the largest impact remnant is the target (Fig.~\ref{fig:5}), and  the metal-silicate mixture in the second-largest impact remnant originated from the projectile. Therefore, our results suggest that mesosiderites may be composed of materials dominantly from its parent body.
        
        Main group pallasites were suggested to have formed in an impact at $\sim1.5-9.5$ Myr after Solar System formation \citep{Windmill2022}, before or during the dissipation of nebular gas when most impacts have low velocities. The dynamo activity recorded by the pallasites \citep{Tarduno2012, Bryson2015, Nichols2021} indicates that the core of the parent body of pallasites was not disrupted after the pallasite-forming impact. Therefore, this impact was likely to be accretional that mixed metal and silicate without disrupting the target's core. Alternatively, a low-velocity hit-and-run impact can produce metal-silicate mixture in the projectile. The buried depth of metal-silicate mixture in the largest remnant after the accretional impacts with $\gamma<0.5$ and metal-silicate mixture in the second-largest remnant after hit-and-run impacts is consistent with the cooling rate of pallasites. On the other hand, the high-$\gamma$ accretional impacts buried most of the metal-silicate mixture at >50 km depth (Fig.~\ref{fig:6}). Therefore, pallasites potentially formed in an accretional impact by a projectile that was less than half the mass of the target or in a low-velocity hit-and-run impact.
    
        Our simulations indicate that some energetic impacts between planetesimals could disrupt the layered structure of differentiated planetesimals, resulting in mixed structures in the impact remnants (Fig.~\ref{fig:1}f). The low-energy accretional impacts between planetesimals could cause stagnation of metal in the silicate parts of planetesimals (Fig.~\ref{fig:1}a and d). These mixed structures could be preserved in asteroids that formed from the remaining planetesimals. It was inferred that the asteroid Psyche may be the metal-rich remnant of a differentiated planetesimal, which contains non-metal components with very low FeO content \citep{Elkins-Tanton2022}. The bulk density of Psyche is similar to those of stony-iron meteorites \citep{Viikinkoski2018}. The silicates in mesosiderites contain too much iron to match the non-metal components of Psyche, while pyroxene pallasites possibly have a parent body that is similar to Psyche \citep{Elkins-Tanton2020}. An impact or multiple impacts, possibly stripping away the mantle of Psyche, could have mechanically mixed the non-metal components into the metal of Psyche.

\section{Conclusions}
    In this work we simulated impacts between planetesimals that span different outcome regimes and analysed the distributions of metal and silicate after impacts. The metal-silicate mixture has been identified and its mass, buried depth, and sources are distinct among impacts with different conditions. Impacts between planetesimals with molten cores generally produce more metal-silicate mixture than impacts between planetesimals with solid cores, indicating that the molten cores could facilitate impact-induced mixing of metal and silicate. Low-energy accretional impacts and high-energy erosive impacts could mix large amounts of metal and silicate in the largest impact remnant. Erosive and hit-and-run impacts could produce metal-silicate mixture in the second-largest impact remnant. Most of the metal-silicate mixture was buried at depth in the impact remnants, which can explain the slow cooling process of stony-iron meteorites. Combined with meteorite records, our results indicate that mesosiderites potentially formed in an erosive impact with a low impact angle ($<45^\circ$), while pallasites potentially formed in an accretional impact by a small projectile ($\gamma<0.5$) or in a low-velocity hit-and-run impact. The mixing of metal and non-metal components on Psyche may also result from impacts.
    
\begin{acknowledgements}
This work was supported by National Natural Science Foundation of China (NSFC) grant (42125303) and the B-type Strategic Priority Program of the Chinese Academy of Sciences (XDB41000000). C.~Burger and C.~M.~Sch{\"a}fer appreciate support by the German Research Foundation - DFG (project 285676328) and support by the High Performance and Cloud Computing Group at the Zentrum für Datenverarbeitung of the University of Tübingen, the state of Baden-Württemberg through bwHPC and the German Research Foundation through grant no INST 37/935-1 FUGG. K. Shuai appreciates support by the China Scholarship Council (202006190255). We acknowledge the computational support from High-Performance Computing Center (HPCC) of Nanjing University and High-Performance Computing Center of Collaborative Innovation Center of Advanced Microstructures. The authors thank the anonymous reviewer for the constructive comments which helped to improve this paper.
\end{acknowledgements}
\bibliographystyle{aa} 
\bibliography{reference.bib}

\begin{appendix}
    \onecolumn
    \section{Tests for ANEOS, high resolutions, purely hydrodynamic simulations, and numbers of particles for Moran's $I$ analysis} 
    \begin{figure}[!htb]
    \centering
    \includegraphics[width=14 cm]{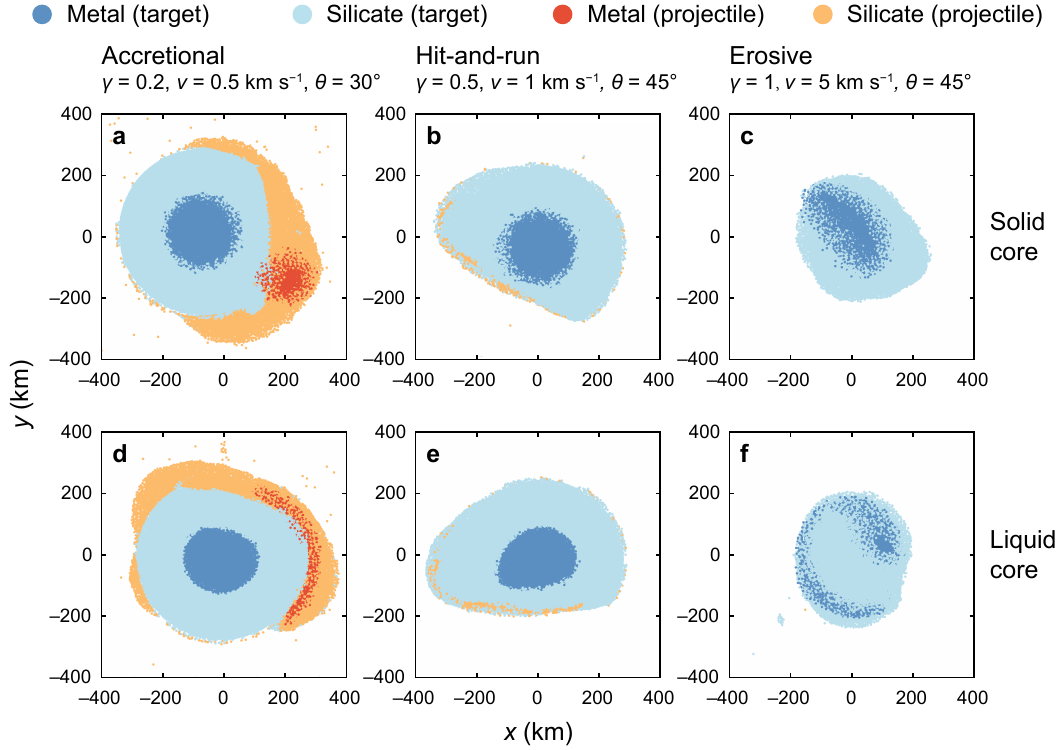}
    \caption{Distribution of metal and silicate at the end of simulations using ANEOS. The cross-sections of the largest impact remnants in different impact regimes are shown; they are in the plane of the projectile trajectory. The colours and the initial conditions of the impacts are the same as in Fig.~\ref{fig:1}.
    }
        \label{fig:A1}
    \end{figure}

    \begin{figure}[!htb]
    \centering
    \includegraphics[width=14 cm]{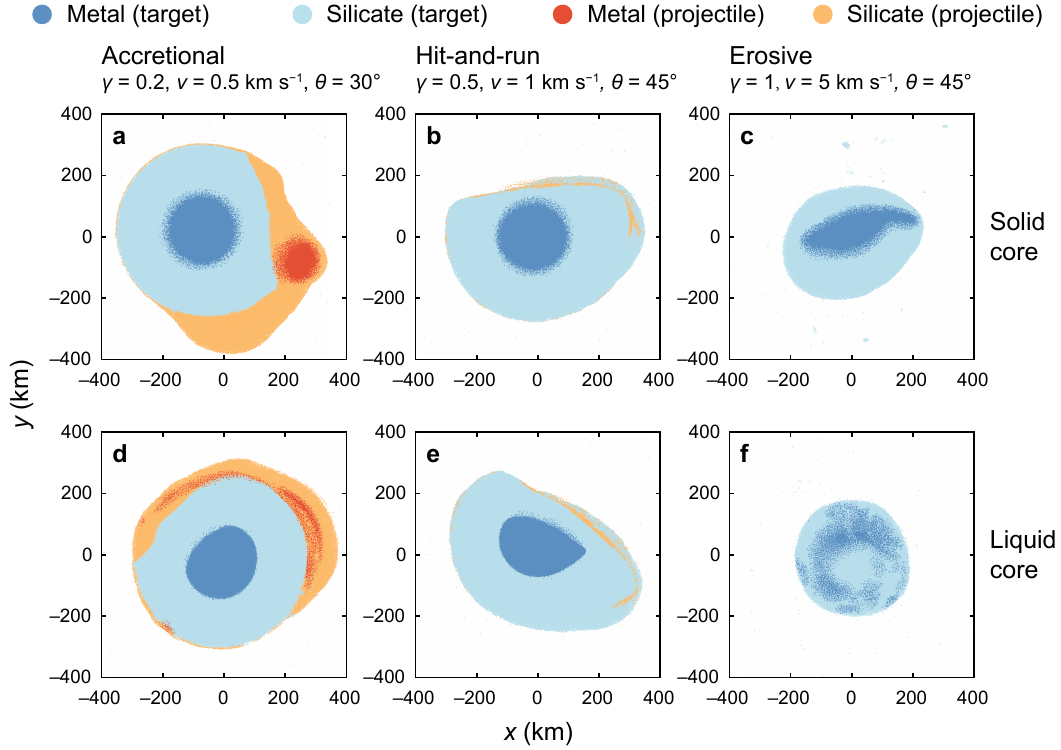}
    \caption{Distribution of metal and silicate at the end of high-resolution simulations. The numbers of particles in the pre-impact bodies are ten times those in the standard-resolution simulations. The cross-sections of the largest impact remnants in different impact regimes are shown; they  are in the plane of the projectile trajectory. The colours and the initial conditions of the impacts are the same as in Fig.~\ref{fig:1}. 
    }
        \label{fig:A2}
    \end{figure}
    \begin{figure}[!htb]
    \centering
    \includegraphics[height=6 cm]{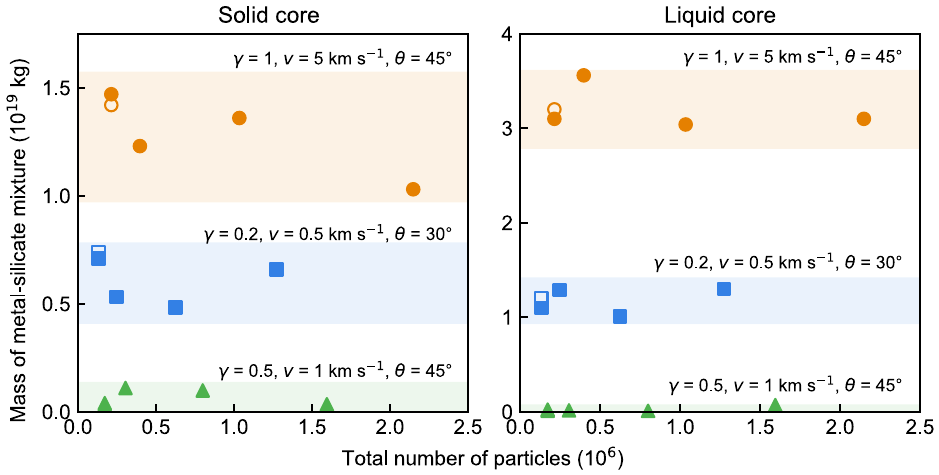}
    \caption{Mass of metal-silicate mixture in the largest impact remnant in simulations with different resolutions and equations of state. For each set of initial conditions, four resolutions were used: 1, 2, 5, and 10 times the standard resolution. The total number of particles in the target and projectile are shown. The solid symbols represent the results of simulations using the Tillotson equation of state and the open symbols represent the results simulated with ANEOS. The shaded bands show $\pm$2$\sigma$ of the masses of metal-silicate mixture for each set of initial conditions.
    }
        \label{fig:A3}
    \end{figure}
    \begin{figure}[!htb]
    \centering
    \includegraphics[width=12 cm]{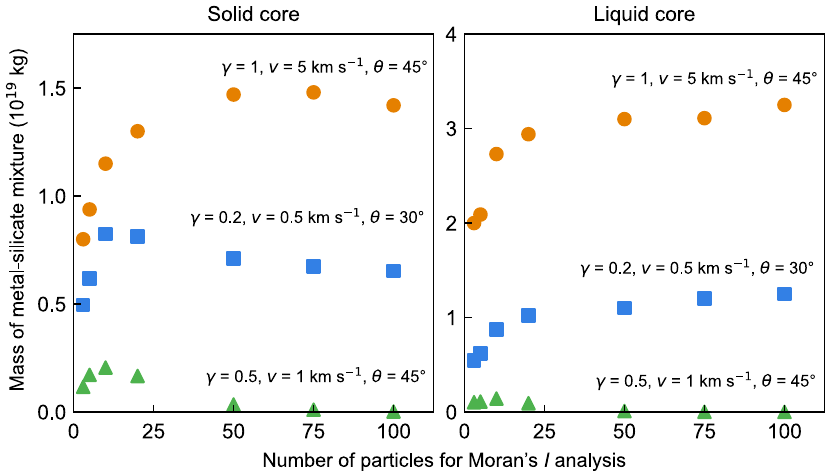}
    \caption{Tests for different numbers of particles for Moran's $I$ analysis ($N$).
    }
        \label{fig:A4}
    \end{figure}
    \begin{figure}[!htb]
    \centering
    \includegraphics[width=14 cm]{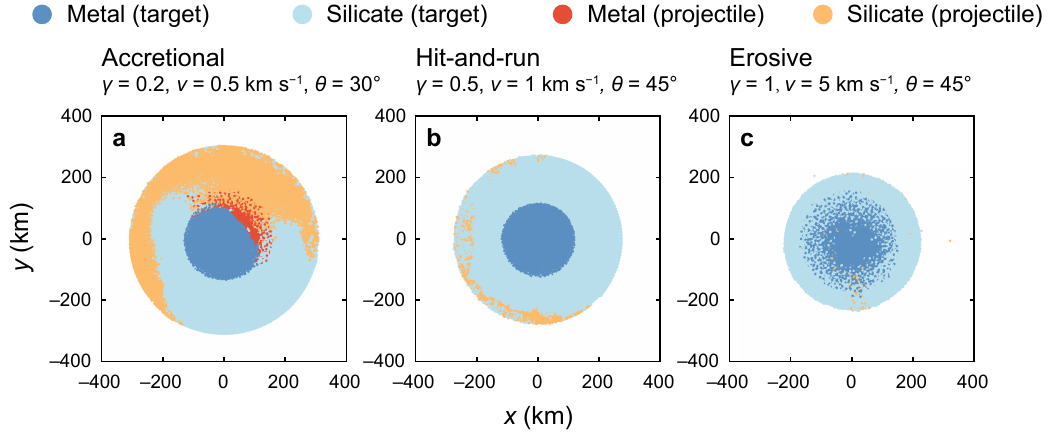}
    \caption{Distribution of metal and silicate at the end of purely hydrodynamic simulations. The cross-sections of the largest impact remnants in different impact regimes are shown; they are in the plane of the projectile trajectory. The colours are the same as in Fig.~\ref{fig:1}.
    }
        \label{fig:A5}
    \end{figure}

    \clearpage
    \section{Material parameters for solid materials}
    \begin{table*}[!htb]
        \footnotesize
        \caption{Material parameters for solid materials.}
        \centering
        \begin{tabular}{l c c c c}
        \hline
         & Solid silicate & Reference & Solid metal & Reference\\
        \hline
        Equation of state & Tillotson basalt & \cite{Benz1999} & Tillotson iron & \cite{Melosh1989} \\
        Shear modulus (GPa) & 22.7 & \cite{Benz1999} & 76 & \cite{Emsenhuber2018} \\
        Shear strength at infinite pressure $Y_{\rm M}$ (GPa) & 3.5 & \cite{Bowling2013} & 2.5 & \cite{Bowling2013} \\
        Cohesion (intact) $c$ (MPa) & 10 & \cite{Bowling2013} & 10 & \cite{Bowling2013} \\
        Coefficient of internal friction (intact) $\mu$ & 1.2 & \cite{Bowling2013} & 2.0 & \cite{Bowling2013} \\
        Cohesion (damaged) $c_{\rm D}$ (Pa) & 10 & \cite{Bowling2013} &  &  \\
        Coefficient of internal friction (damaged) $\mu_{\rm D}$ & 0.6 & \cite{Bowling2013} &  &  \\
        Weibull parameter $m$ & 16 & \cite{Nakamura2007} & & \\
        Weibull parameter $k$ (m$^{-3}$) & 10$^{61}$ & \cite{Nakamura2007} & & \\
        Melt energy $E_{\rm melt}$ (MJ kg$^{-1}$) & 8.7 & \cite{Quintana2015} & 6.7 & \cite{Quintana2015} \\
        \hline
        \end{tabular}
        \label{table:B1}
    \end{table*}
    
    \begin{multicols}{2}
    \section{Criteria of impact regime classification\label{appendix:c}}
    Different criteria were used to define impact regimes in previous studies. The boundary between accretional and hit-and-run regimes in this study is generally consistent with that in the previous study \citep{Leinhardt2012}, with most accretional impacts having an impact angle lower than the critical impact angle: $\sin\theta<R_{\rm t}/(R_{\rm t}+R_{\rm p})$, where $R_{\rm t}$ and $R_{\rm p}$ are the radii of the target and the projectile. The exceptions are a few accretional impacts with low $v$ and high $\theta$, in which the impacting bodies merged, and a few hit-and-run impacts with high $v$ and low $\theta$, in which the target accreted a small amount of mass from the projectile. These exceptions are consistent with a recent study that also found hit-and-run impacts with $\theta$ lower than the critical impact angle, which suggested that the transition between grazing and non-grazing regimes is not a hard boundary based on the critical impact angle \citep{Cambioni2019}.
    
\begin{figure*}[!htb]
    \sidecaption
    \includegraphics[width=12cm]{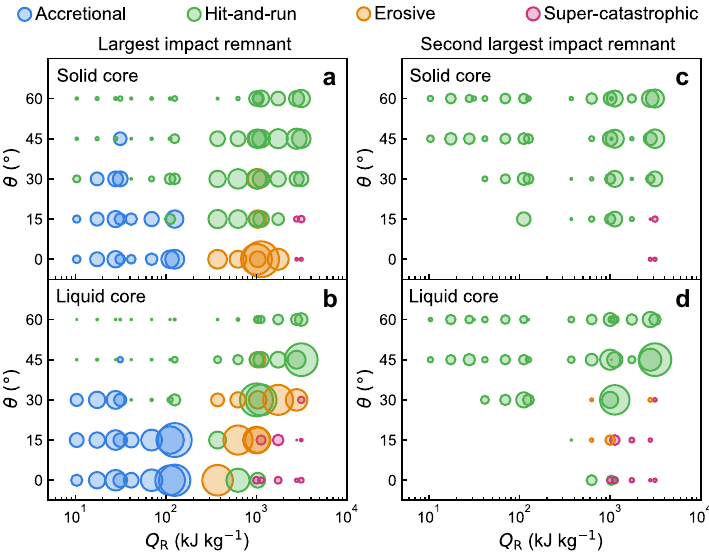}
    \caption{Mass of metal-silicate mixture in the largest and the second-largest impact remnants. Each circle corresponds to one impact simulation. The sizes of the circles are proportional to the mass of metal-silicate mixture produced in the impacts. The impacts with higher impact velocities and higher projectile-to-target ratios have higher impact energies ($Q_{\rm R}$). The outcome regimes of impacts are classified according to the mass of the second-largest impact remnant.
    }
        \label{fig:C1}
    \end{figure*}

    In several hit-and-run impacts, $M_{\rm l}$ is slightly less than $M_{\rm t}$. These impacts were classified as erosive hit-and-run impacts in the previous study \citep{Leinhardt2012}. Because the target's core was almost unaffected and only a small amount of metal-silicate mixture formed in these impacts, we combined this erosive hit-and-run regime with the hit-and-run regime. On the other hand, the mass of the second-largest impact remnant ($M_{\rm sl}$) has been used to define impact regimes, with hit-and-run impacts having $M_{\rm sl} > 0.1 M_{\rm p}$ \citep{Emsenhuber2020, Asphaug2021}. We applied this criterion to our results. The impact regimes defined based on $M_{\rm sl}$ (Fig.~\ref{fig:C1}) are different from those defined by accretion efficiency (Fig.~\ref{fig:4}). Some impacts capable of producing high masses of metal-silicate mixture in the largest impact remnant can be classified as hit-and-run impacts according to $M_{\rm sl}$. Because our criterion based on accretion efficiency can  allow us to clearly discern which impacts can produce high masses of metal-silicate mixture in the largest impact remnant (Fig.~\ref{fig:4}), we mainly discuss our results according to this criterion in the main text. On the other hand, if the criterion based on $M_{\rm sl}$ is used, almost all impacts that can mix metal and silicate in the second-largest impact remnant are defined as hit-and-run impacts (Fig.~\ref{fig:C1}c and d). Therefore, as long as the projectile survived in the impact, the metal and silicate in it can be mixed, suggesting that impacts are transformative for the projectile.
    \end{multicols}
\end{appendix}

\end{document}